\documentclass[letter,twocolumn]{jpsj3}
\usepackage{txfonts}
\usepackage{amssymb}
\usepackage{amsmath}
\usepackage{bm}
\usepackage{graphicx}
\usepackage{dcolumn}
\usepackage{longtable}
\usepackage{color}

\title{Emergence of Frustrated Short-Range Order above Long-Range Order in the $S=1/2$ Kagome Antiferromagnet CaCu$_3$(OD)$_6$Cl$_2\cdot0.6$D$_2$O}

\author{Yoshihiko~Ihara$^1$\thanks{yihara@phys.sci.hokudai.ac.jp}, Kazuki~Matsui$^2$, Yoshimitsu~Kohama$^2$, Sven~Luther$^{3,4}$, Daryna~Opherden$^3$, Jochen~Wosnitza$^{3,4}$, Hannes K\"{u}hne$^3$, Hiroyuki K. Yoshida$^1$}
\inst{$^1$Department of Physics, Faculty of Science, Hokkaido University, Sapporo 060-0810, Japan \\
$^2$The Institute for Solid State Physics, University of Tokyo, Kashiwa, Chiba 277-8581, Japan\\
$^3$Hochfeld-Magnetlabor Dresden (HLD-EMFL) and W\"{u}rzburg-Dresden Cluster of Excellence ct.qmat, Helmholtz-Zentrum Dresden-Rossendorf, 01328 Dresden, Germany\\
$^4$Institut f\"{u}r Festk\"{o}rper-und Materialphysik, TU Dresden, 01062 Dresden, Germany} 

\abst{
We report on the low-energy dynamics in the kagome antiferromagnet CaCu$_3$(OD)$_6$Cl$_2\cdot0.6$D$_2$O (Ca-kapellasite) 
as studied by use of $^2$D-NMR measurements. 
Previous $^{35}$Cl-NMR measurements revealed that the nuclear spin-lattice relaxation rate ($1/T_1$) shows two peaks at temperatures, 
$T^{\ast} = 7.2$ K and $T_s \simeq 25$ K. 
While the low-temperature peak at $T^{\ast}$ is ascribed to the critical fluctuations near the long-range magnetic ordering, 
the origin of the high-temperature peak has not been fully understood. 
From the $1/T_1$ measurements on the D sites at the OD groups (D$_{\rm OD}$), we find no peak at $T_s$, 
evidencing that the high-temperature peak is not related to the molecular dynamics of the OD groups. 
We discuss the possibility of a frustration-induced short-range ordered state below $T_s$ before the long-range order is stabilized by the Dzyaloshinskii-Moriya interaction. 
We also observed static internal fields at the D$_{\rm OD}$ site in the long-range ordered state below $T^{\ast}$, and confirm the 
previously proposed negative-chirality $q=0$ magnetic structure. }

\begin{document}
\maketitle

In a geometrically frustrated magnet competing magnetic interactions prevent long-range magnetic order, 
leading to a disordered spin state with fluctuations even at very low temperatures. 
Intriguing effects appear when quantum-mechanical zero-point fluctuations are introduced to such a disordered state. 
The most intensively studied example is a quantum spin liquid \cite{balents-Nature464}, which is suggested to appear in a Heisenberg kagome antiferromagnet (KAFM). 
Here, significant spin fluctuations are expected because of a small number of nearest neighbor spins $(Z=4)$. 
In a real material, the quantum spin liquid has been experimentally identified in the mineral herbertsmithite \cite{shores-JACS127, norman-RMP88, imai-PRL100, oraliu-PRL100},
although the determination of its character, namely gapped topological $Z_2$ state \cite{yan-Science332} or gapless $U(1)$-Dirac state \cite{ran-PRL98}, 
is still left for further study. \cite{fu-Science350, jeong-PRL107, khuntia-NatPhys16}
Another family of model systems for the Heisenberg KAFM are the kapellasite minerals, in which $S=1/2$ Cu$^{2+}$ spins construct a perfect kagome network. 
The magnetism in kapellasite is described on the basis of a $J_1-J_2-J_d$ model \cite{iqbal-PRB92} because of a small but finite long-range interaction across the kagome hexagon, 
which is assisted by the non-magnetic ions (Zn, Mg, Cd, etc.) intercalated at the center of the hexagon. \cite{fak-PRL109, okuma-natcommun10, kermarrec-PRB90, boldrin-PRB91, sun-JMCC4, zorko-PRB99}

\begin{figure}
\begin{center}
\includegraphics[width=8cm]{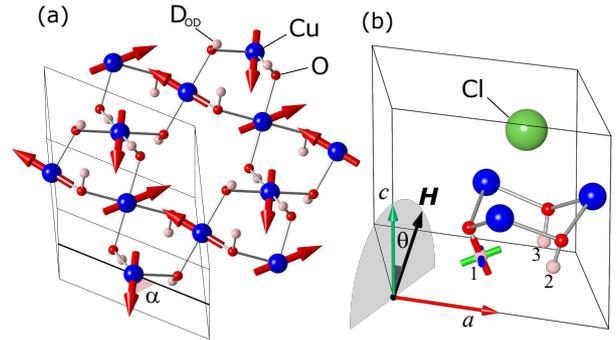}
\end{center}
\caption{(Color online)
(a) Kagome network of Cu$^{2+}$ spins. 
The arrows on the Cu sites indicate the spin direction in the negative-chirality $q=0$ structure. 
The thin black lines represent a unit cell. 
The freedom of global rotation, inherent in $120^{\circ}$ structures is characterized by the angle $\alpha$ between the spin direction and $a$ axis. 
(b) Closer view around the D$_{\rm OD}$ sites. 
The principal axes of the electric field gradient (EFG) are represented by thick bars on the site 1. 
The largest EFG orients to the OD bond direction (red bar).
In this study, the external magnetic field is rotated around the $a$ axis and its direction is identified by the angle $\theta$ from the $c$ axis.
}
\label{fig1}
\end{figure}

Among the series of kapellasites, Ca-kapellasite [CaCu$_3$(OD)$_6$Cl$_2\cdot0.6$D$_2$O] is special 
because of its negative $J_1$, absence of Ca/Cu sites mixing, and availability of single crystals. \cite{yoshida-JPSJ86, sun-PCM43}
From our previous studies we unveiled that the magnetic ordering at $T^{\ast}=7.2$ K is caused by the antisymmetric Dzyaloshinskii-Moriya (DM) interaction, 
which selects a chirality ordered $q=0$ magnetic structure. [Fig.~\ref{fig1}(a)] \cite{ihara-PRR2} 
This long-range ordered magnetic structure is confirmed by neutron diffraction. \cite{iida-PRB101}
From our previous nuclear spin-lattice relaxation rate ($1/T_1$) measurements\cite{ihara-PRR2}, we also found persistent low-energy magnetic fluctuations in the spin component perpendicular to the kagome plane, 
while those within the kagome plane are explained simply by the dispersive magnon excitations from the $q=0$ ground state. 
In contrast to the detailed study for the magnetic fluctuations deep in the ordered state, 
the peak in $1/T_1$ found at a temperature higher than $T^{\ast}$ has not been sufficiently addressed so far. 
In Ca-kapellasite, the temperature dependence of $1/T_1$ shows two peaks at $T^{\ast}$ and $T_s \simeq 25$ K. \cite{ihara-PRB96}
The former is caused by the critical slowing down at the long-range magnetic phase transition. 
A magnetic origin of the latter was previously suggested, namely the growth of short-range spin correlations. 
This was motivated by the observation of broad maxima of the magnetic susceptibility and heat capacity at the corresponding temperature. \cite{yoshida-JPSJ86}
However, since the peak temperature increases in high magnetic fields, corresponding to a higher resonance frequency, 
molecular dynamics of the OD groups as a possible origin of the $1/T_1$ peak need to be investigated. \cite{opherden-arxiv}
In fact, a similar two-peak behavior was observed in herbertsmithite, \cite{imai-PRL100,jeong-PRL107}
where the peak at higher temperature is ascribed to lattice vibrations as concluded from the frequency and site dependence of the peak temperatures. 
Previous $^{35}$Cl-NMR measurements did not allow to identify the origin of the high-temperature peak, 
because the $^{35}$Cl nuclei with spin $I =3/2$ are located at a position, where both magnetic and electric fluctuations can contribute to the nuclear-spin relaxation process.

In this study, we investigate the origin of the high-temperature peak utilizing $^2$D-NMR measurements. 
We used a deuterated single crystal because $^2$D nuclei with spin $I=1$ can couple to the electric fluctuations through the electric quadrupole interaction, 
and the gyromagnetic ratio for $^2$D is close to that of $^{35}$Cl, which allows us to measure $1/T_{1}$ at a comparable field and frequency. 
Since the molecular-dynamical effects originate in the OD group with a small mass, 
large features should be detected in the $1/T_1$ measurements by the $^2$D-NMR signal of the OD groups. 
We also measured the $^2$D-NMR spectra in the ordered state to confirm the previously suggested $q=0$ magnetic structure with negative-chirality spin configuration. 
In our previous study, the spin chirality was clearly identified from the direction of the internal fields at the Cl site on a trigonal axis. \cite{ihara-PRR2}
More complicated spectra are expected for D placed on less symmetric positions close to the mirror plane. 
Thus, we simulate the internal fields at the D sites on the basis of the negative-chirality $q=0$ structure, 
and compare the results with the experimentally obtained spectrum. 

The deuterated single crystals were prepared by a hydrothermal reaction using a 2-zone tube furnace. 
Chemical reagents of anhydrous calcium chloride and copper oxide with deuterium oxide $(99.8 \% )$ were sealed into a quartz tube, 
and heated for two weeks keeping a temperature gradient of 50 $^{\circ}$C with highest temperature of 220~$^{\circ}$C. 
By that, blue hexagonal plates were obtained. \cite{yoshida-JPSJ86}
The mass of the crystal used for the NMR measurements is $\sim 7$ mg. 
$1/T_1$ was measured using the conventional saturation-recovery method. 
Above $T^{\ast}$, the recovery profile of the nuclear magnetization fits well to the theoretical curve for a nuclear spin $I=1$. 
We measured $1/T_1$ at the D as well as at the Cl sites for the same deuterated crystal to exclude that the deuteration affects the intrinsic magnetic properties. 
The NMR spectra in the ordered state were measured by sweeping the field at a constant frequency of $32.31$ MHz due to a broad spectral width. 
The spectra at room temperature were considerably sharper than those in the ordered state. 
Thus, we measured FFT spectra at a fixed magnetic field of $13.0266$ T.
The orientation of the external magnetic field was tuned by one-axis rotator, 
with which the sample can be rotated around the crystalline $a$ axis. 
The field direction is given by the angle $\theta$ measured from the $c$ axis as shown in Fig.~\ref{fig1} (b).

\begin{figure}
\begin{center}
\includegraphics[width=7cm]{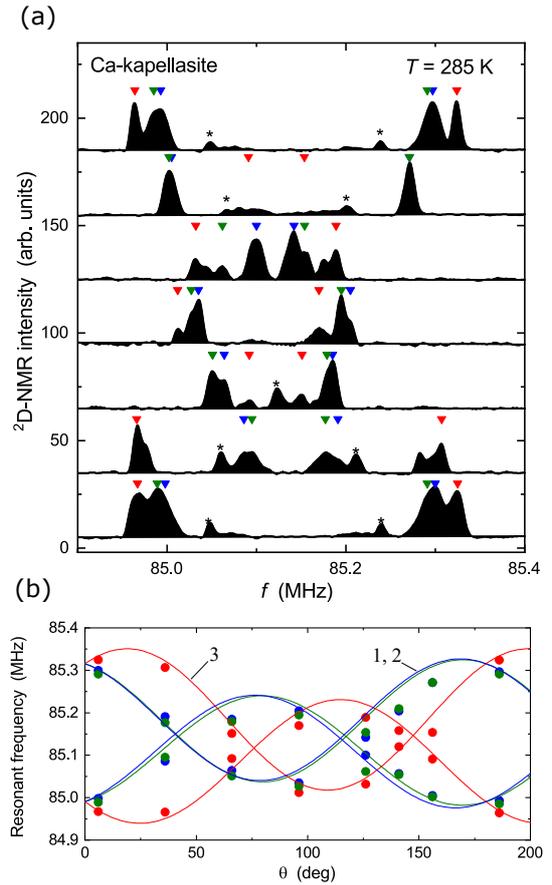}
\end{center}
\caption{(Color online)
(a) $^{2}$D-NMR spectra for various orientations of the external magnetic field with respect to the $c$ axis. 
The spectra are offset by the angle $\theta$ for clarity. 
The peak positions of the different D$_{\rm OD}$ sites are marked by triangles. 
We detected extra peaks marked by asterisks, which were assigned to D$_{\rm D2O}$ sites. 
(b) Peak positions plotted as a function of $\theta$. 
The solid lines are the result of a numerical simulation. 
Three branches of line pairs originate from three D$_{\rm OD}$ sites. 
The site numbers 1, 2 and 3 correspond to those displayed in Fig.~\ref{fig1} (b). 
}
\label{fig2}
\end{figure}

To understand the microscopic dynamics around $T_s$ and the magnetic structure below $T^{\ast}$, 
the $^2$D-NMR peaks need to be assigned to the two cyrstallographically different D sites in Ca-kapellasite, 
which are the D sites at the OD group (D$_{\rm OD}$) and those at the partially occupied crystal water position (D$_{\rm D2O}$). 
We, therefore, measured the angular dependence of the NMR spectra at room temperature as shown in Fig.~\ref{fig2} (a). 
As the peak positions are determined by the local environment around the target D sites, 
we can identify the D$_{\rm OD}$ sites from the angular dependence. 
For $^2$D nuclei with a nuclear spin of $I=1$, two NMR peaks from transitions between $m_z=-1 \leftrightarrow 0$ and $m_z=0 \leftrightarrow +1$ are observed. 
In addition, the three crystallographically equivalent D$_{\rm OD}$ sites reproduced by $C_3$ operation become magnetically non-equivalent for an arbitrary field direction. 
As a result, 6 NMR peaks from the D$_{\rm OD}$ sites were detected as marked by downward triangles in Fig.~\ref{fig2} (a). 
Additional peaks indicated by asterisks are assigned to the signal from D$_{\rm D2O}$.
The NMR spectrum from D$_{\rm D2O}$ is broad because of disorder in the D$_2$O orientation, and,
thus, we cannot extract any clear angular dependence at room temperature.

To confirm the above peak assignment more quantitatively, we calculated the angular dependence of the peak frequencies. 
The $^2$D-NMR frequencies are determined both by the magnetic and electric quadrupole interactions, 
as represented by the following Hamiltonian: 
\begin{align}
\mathcal{H} &= -\gamma \hbar \left[1+K(\theta) \right] \bm{H}_0 \cdot \bm{I}  \notag \\
&+ \frac{e^2qQ}{4I(2I-1)}\left[ \left( 3I_z^2-\bm{I}^2 \right) + \frac{1}{2}\eta \left( I_{+}^2+I_{-}^2 \right) \right].
\end{align}
Here, $Q, eq,$ and $\eta$ are the electric quadrupole moment, principal ($z$) component of the electric field gradient (EFG) at the D site, 
and the asymmetry parameter, respectively. 
$K(\theta)$ is the anisotropic Knight shift mostly originating from the dipole fields of the Cu$^{2+}$ moments. \cite{sup}
The EFG tensors for the D sites were calculated by using a point-charge model. 
The result shows that the main principal axis of the EFG orients along the OD bond direction as shown in Fig.~\ref{fig1} (b) by the red bar on the D$_{\rm OD}$ site-1, 
and the NQR frequency $\nu_{Q}$ and $\eta$ are $\nu_{Q}=682$ kHz and $\eta=0.034$. \cite{sup}
We also calculated the anisotropic Knight shift caused by dipole interaction from the paramagnetic Cu$^{2+}$ spins. 
The result is 
\begin{align}
A_{\rm dip}= \left(
\begin{array}{rrr}
-78.2 & -69.0 & 37.0 \\
-69.0 & 1.8 & 25.3 \\
37.0 & 25.3 & 76.4 
\end{array}
\right) \; {\rm mT}/\mu_B , 
\end{align} 
with the directions of the unit vectors along the $a$, $b^{\ast}$ and $c$ axes. 
With these parameters, we calculated the angular dependence of the peak positions and show the result as solid lines in Fig.~\ref{fig2} (b). 
Three branches indicated by numbers correspond to the three D$_{\rm OD}$ sites in Fig.~\ref{fig1} (b).
We introduced a reduction factor of $0.6$ for $\nu_Q$ to describe the experimental result, which most probably stems from the covalent character of the OD bond. 
The resulting angular dependence consistently explains the experimentally obtained peak positions, and thus evidences the validity of our site assignment.

\begin{figure}
\begin{center}
\includegraphics[width=7cm]{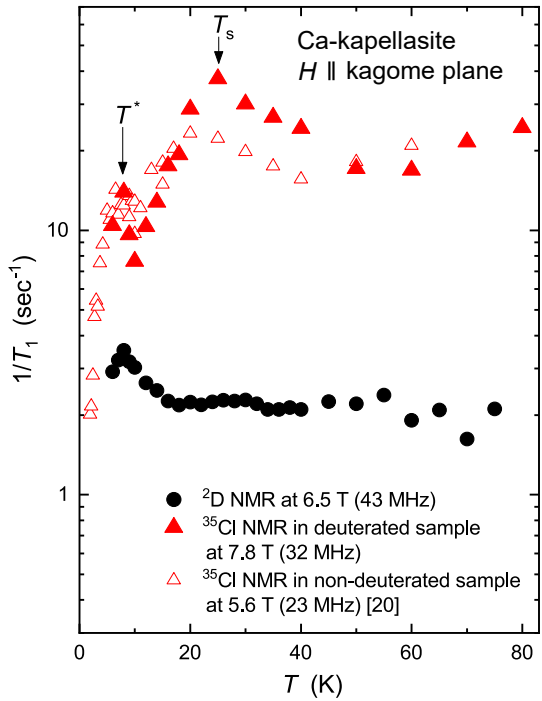}
\end{center}
\caption{(Color online)
Temperature dependence of $1/T_1$ measured at the D (circle) and Cl (triangle) sites. 
The peak at $T_s$ is observed at the Cl site also in the deuterated sample (filled triangles), 
while this peak is absent at the D site. 
In contrast, the peak at $T^{\ast}$ is observed for both sites. 
Results from a previous study for a not deuterated sample (empty triangles) are shown for comparison. \cite{ihara-PRR2}
}
\label{fig3}
\end{figure}

Figure \ref{fig3} shows the temperature dependence of $1/T_1$ measured for fields parallel to the kagome plane ($\theta = 90^{\circ}$). 
We determined $1/T_1$ at both the $^2$D and $^{35}$Cl sites for the same deuterated single crystal. 
The external magnetic fields were $6.5$ T and $7.8$ T for the $^2$D- and $^{35}$Cl-NMR measurements, 
which correspond to NMR frequencies of approximately $43$ MHz and $32$ MHz, respectively. 
In addition to the present results, we show previous results measured at $23$ MHz by $^{35}$Cl NMR for a non-deuterated sample for comparison. \cite{ihara-PRR2}
The reproducible temperature dependence at the Cl sites confirms 
that the intrinsic magnetic properties are not significantly modified by deuteration. 
In contrast, $1/T_1$ at the D$_{\rm OD}$ sites shows only one peak at $T^{\ast}$ without any detectable anomaly around $T_s$. 
As a possible molecular motion of the OD group should be most sensitively detected by the on-site $^2$D-NMR measurement, 
the absence of an anomaly around $T_s$ excludes the possibility of molecular dynamics as the origin of the peak at $T_s$ in $1/T_1$ for the Cl site.

Therefore, this strongly suggests that the peak at $T_s$ is of a magnetic origin associated with the formation of short-range correlations. 
The broad hump in the magnetic susceptibility and specific heat at the corresponding temperature indicates that the spins start 
to release entropy by the formation of antiferromagnetic spin configurations for short time duration and distance. 
The absence of a peak in $1/T_1$ at $T_s$ at D$_{\rm OD}$ sites originates in the cancellation of the internal fields at D$_{\rm OD}$ sites 
that are nearly on the mirror plane for neighboring Cu spins that form anti-parallel spin configurations. 
In contrast, the peak of $1/T_{1}$ at $T_s$ is observed at the Cl sites on the trigonal axis 
because geometrical frustration prohibits anti-parallel spin configurations for all three Cu spins equally coupled to a given Cl nuclear spin,  
and thus, generates purely dynamical internal fields at the Cl sites. 
In the temperature range between $T_s$ and $T^{\ast}$, the nearest-neighbor interaction $J_1$ dominates the spin correlation, 
as the energy scales of exchange interactions determined from bulk susceptibility measurements are $J_1 = 52.2$ K, $J_2=-6.9$~K, and $J_d = 11.9$ K. \cite{yoshida-JPSJ86}
Since the spin-singlet pairs formed neighboring Cu spins with an exchange coupling $J_1$ are highly degenerate because of the geometrical frustration, 
a short-range resonating valence bond state is expected to appear at low temperatures by the introduction of quantum fluctuations. \cite{zeng-PRB51}
In fact, the magnetic state below $T_s$ can be understood in terms of a spin liquid as suggested from thermal Hall-effect experiments. \cite{doki-PRL121}
When the DM interaction starts affecting the spin correlations at lower temperatures approaching $T^{\ast}$, 
the short-range ordered spin state is finally superseded by a long-range ordered magnetic state.  
The critical fluctuations associated with the long-range phase transition appear as a peak in $1/T_1$, 
since the $120^{\circ}$ spin structure creates a finite internal field both at the D$_{\rm OD}$ and Cl sites.

\begin{figure}
\begin{center}
\includegraphics[width=8cm]{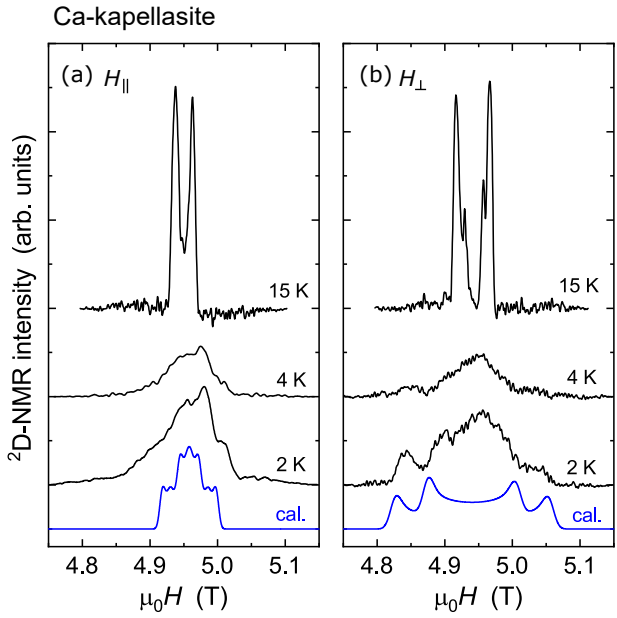}
\end{center}
\caption{(Color online)
$^2$D-NMR spectra measured at low temperatures near $T^{\ast} = 7.2$ K. 
Below $T^{\ast}$, the spectral broadening is caused by the static internal fields for both field directions 
(a) parallel ($H_{\parallel}$) and (b) perpendicular ($H_{\perp}$) to the kagome plane. 
The simulated spectra for a negative-chirality $q=0$ structure are shown at the bottom. 
The spectral width is consistent with the experimental data, while additional features are observed in the experimental spectra, 
which is caused by a non-uniform $\alpha$ distribution (see text).
}
\label{fig4}
\end{figure}

Now, we focus on the $^{2}$D-NMR spectra in the ordered state, 
where the $^2$D local fields are composed of both the direct dipole and transferred hyperfine fields from the ordered Cu spins. 
The $^2$D-NMR spectra for fields parallel ($H_{\parallel}$) and perpendicular ($H_{\perp}$) to the kagome plane are broadened below $T^{\ast}$ as shown in Fig.~\ref{fig4}. 
We found that the spectrum for $H_{\perp}$ is significantly broader than that for $H_{\parallel}$, which is opposite to the result of the $^{35}$Cl-NMR measurement, 
where the spectral width is very small for $H_{\perp}$. \cite{ihara-PRR2}
To understand the $^2$D-NMR spectra, we first calculated the dipole fields at the D$_{\rm OD}$ sites based on the negative-chirality $q =0$ magnetic structure. 
The local field varies with the global rotation of ordered moments, which is allowed for the coplanar $120^{\circ}$ structure. 
The global rotation is characterized by the angle $\alpha$ between the direction of a spin at one Cu site and the $a$ axis [Fig.~\ref{fig1} (a)]. 
We summed up the dipole fields from magnetic moments of $1 \mu_B$ located at the Cu sites within a radius of about $30$ \AA \; from the target D site. 
The calculated dipole field is written as $B_{\rm dip}^{\xi}(\alpha) = B^{\xi}\sin(\alpha +\phi_{\xi})$, $(\xi = x,y,z)$ with 
\begin{align}
\bm{B} &= \left[ -47 ~{\rm mT}, -28~{\rm mT}, 155~{\rm mT} \right] , \\ 
\bm{\phi} &= \left[ 1^{\circ}, 8^{\circ}, -0.2^{\circ} \right]. \label{eq:phi}
\end{align} 
$B^z$ larger than $B^{x,y}$ is compatible with the experimental results. 
If we assume a positive-chirality spin configuration, the resulting field $\bm{B}$ has nearly the same value of about $100$ mT for all components, 
which does not agree with the angular dependence of the $^2$D-NMR spectra. 

Next, we estimate the contribution from hyperfine fields. 
The total coupling constant between the electronic and $^2$D nuclear spins was determined as $A_{\rm total} = 66(4)$ mT/$\mu_B$ 
by the linear relationship between the bulk susceptibility $\chi$ and Knight shift $K$ in the paramagnetic state. \cite{sup}
By subtracting the numerically computed dipole coupling along the $c$ direction, $A_{\rm dip} = 76.4$ mT/$\mu_B$, 
we obtain an isotropic hyperfine coupling constant $\overline{A}_{\rm hf} =-10$ mT/$\mu_B$. 
Assuming an equal coupling strength for two neighboring Cu spins, we estimate the hyperfine coupling constant for one Cu spin as $A_{\rm hf} = -5$ mT/$\mu_B$. 
Since this hyperfine field is much smaller than the dipole fields estimated above, we can neglect this contribution. 

We simulated the NMR spectrum in the ordered state by assuming a uniform distribution of $\alpha$ and a reduced moment of $0.6 \mu_B$ 
and show the results as the lowest (blue) curve in Fig.~\ref{fig4}. 
The spectral widths for both $H_{\parallel}$ and $H_{\perp}$ are in very good agreement with the experimental spectra, 
which strongly supports the negative-chirality $q=0$ structure. 
The reduced ordered moment suggests remaining quantum fluctuations in the ordered state, 
which may be the origin of the $T$-linear term in the heat capacity and $1/T_{1}$. \cite{yoshida-JPSJ86, ihara-PRB96} 
We note that some features of the spectrum, such as the peak at approximately $4.85$ T in $H_{\perp}$, are explained by the simulation. 
However, the experimental spectrum shows a large intensity near the center of the spectrum, which is absent in the simulation. 
We suggest that this discrepancy is caused by the non-uniform distribution of $\alpha$. 
As a large intensity was observed near the center of the spectrum, we suggest that the $\alpha$ distribution has a large weight near $\alpha \simeq 0$, 
at which the internal field becomes small as can be seen from the dipole field $B_{\rm dip}^{\xi}(\alpha)$ given above. 
The non-uniform distribution of $\alpha$ was also observed in the $^{35}$Cl-NMR spectra as the structureless broad spectrum for $H_{\parallel}$, \cite{ihara-PRR2}
and by the field dependence of the thermal Hall effects. \cite{akazawa-arxiv}

To summarize, we performed $^2$D-NMR measurements for the D$_{\rm OD}$ site and 
confirmed the absence of a peak in $1/T_1$ around $T_s$, which was previously detected by the $^{35}$Cl-NMR measurements. 
This result excludes the possibility of a molecular motion of the OD group as the origin of this peak in $1/T_1$. 
We suggest that the peak originates in the formation of antiferromagnetic spin configurations on short distance and time scales. 
The anti-parallel spin configuration, favored by the dominant nearest-neighbor interaction $J_1$, cannot extend to long-range order 
because of geometrical frustration effects. 
Thus, spin liquid appears below $T_s$ until the negative-chirality $q=0$ structure is selected by the DM interaction at $T^{\ast}$. 

\begin{acknowledgements}
We would like to acknowledge J. Ohara for fruitful discussions. 
This study was partly supported by the JSPS Grant-in-Aid for Scientific Research (Grant Nos. 15K17686, 18K03529, 18H01163 and 19H01832).
We acknowledge support from the Deutsche Forschungsgemeinschaft
(DFG) through SFB 1143 and the W\"{u}rzburg-Dresden Cluster of Excellence on
Complexity and Topology in Quantum Matter--$ct.qmat$ (EXC 2147, Project
No. 390858490), as well as by the HLD at HZDR, a member of the European
Magnetic Field Laboratory (EMFL).
\end{acknowledgements}

\end{document}